\documentclass[
 ,draft            
  ]
  {aipproc}

\layoutstyle{6x9}

\usepackage{epsfig}
\usepackage{graphics}

\begin{document}

\title{Gravitational Wave Sources from  New Physics}
\author{Craig J. Hogan}{
address= {Astronomy and Physics Departments, University of Washington,Seattle, Washington 98195-1580}
}
\begin{abstract}
Forthcoming advances in  direct gravitational wave detection from kilohertz to nanohertz frequencies have unique capabilities to detect signatures from or set meaningful constraints on a wide range of new cosmological phenomena and new fundamental physics. 
A brief survey is presented of the post-inflationary  gravitational radiation backgrounds predicted in  cosmologies that include intense new classical sources such as first-order phase transitions, late-ending inflation, and dynamically active mesoscopic extra dimensions.  LISA will provide the most sensitive direct probes of such phenomena  near  TeV energies or Terascale. LISA will also deeply probe the broadband background, and possibly bursts, from  loops of cosmic superstrings predicted to form in current models of brane inflation.
\end{abstract}
\maketitle

\section{introduction}
In recent years, cosmology has converged on a ``concordance'' model that continues to meet with remarkable success wherever new observational tests are introduced. The direct data span a large dynamic range of structural scales, including power spectrum probes over about a factor of $10^4$ in linear comoving scale.  The microwave background spectrum  and the products of cosmological nucleosynthesis confirm basic model assumptions and microphysics back to the first fraction of a second of the Big Bang.  

An important part of the standard scenario is inflation.  The generic inflationary  predictions of  near global flatness and nearly scale-invariant, Gaussian fluctuations are well confirmed by the data. We thus seem to have some direct information, on very large comoving scales, dating from some very early time, prior to the Terascale (or TeV energies), when the Hubble length was less than about a millimeter and the Hubble time less than a few picoseconds.

At the same time, advances in string theory have opened up a wide range of new possibilities for the initial conditions and the behavior of the universe at early times.  Seven extra dimensions seem mandatory for consistency; they must be  compactified, or gravity must be localized by some other means;  depending on the order of events, these possibilities may require extensions to the usual framework of inflation based on field theory in 3+1 dimensions. Initial conditions, and even apparently fundamental parameters such as the shapes of effective potentials in the field theory Lagrangian,   may be strongly shaped by anthropic selection from a string-theory ``landscape'', changing  the character of arguments about what is natural for inflationary scenarios.  Branes of various dimensionalities  introduce new behavior not present  in a system of fields in 3D space, including  new possibilities for inflation itself, for the generation of fluctuations, and for new kinds of macroscopic defects such as cosmic superstrings.  

Such a large range of new possibilities can be introduced because the observational tests of concordance cosmology cannot directly probe a very large swathe of early cosmic history. Viewed in terms of logarithmic scales in energy space or cosmic scale factor, the unexplored region of our past light cone greatly  exceeds the explored region. For example, information from particle content only survives from the decoupling of the relevant species; for most dark matter candidates, this is well below a GeV in temperature. The baryon content is a presumed relic of the pre-TeV period, which sets the only firm lower limit on the end of inflation. And while it is likely true that the largest-scale perturbations date from very early,  many (possibly as many as about 27)  orders of magnitude of expansion before the end of  inflation, the direct data on these cover only  a relatively brief period and a few orders of magnitude of comoving scale. We have almost no direct data on what happened during about 30 orders of magnitude of cosmic expansion; for almost all of this period, we do not even have solid theoretical arguments to believe that we understand the  dimensionality of space.

Advances in detection of high frequency gravitational waves are starting to provide a way to explore  observationally this very large range of scales, epochs and phenomena not previously accessible. A new generation of ground-based detectors operating in the  sub-kilohertz band\cite{LIGOURL,VIRGOURL}, such as LIGO, GEO, and VIRGO, are now collecting data at their design sensitivity, and will be significantly upgraded in the years ahead.  The space interferometer\cite{LISAURL}  LISA, operating in the millihertz band, will have its core technology proven in a test flight in 2009 and will itself fly in about ten years. Pulsar surveys are providing larger samples of clean pulsars for timing-based detection at frequencies of inverse years to decades\cite{Kaspi:1994hp,Thorsett:1996dr,McHugh:1996hd,Lommen:2001ax,Lommen:2002je}. Together, these techniques are maturing to the point where we will have cosmologically sensitive information over a range of frequencies from  kilohertz to nanohertz. Since gravitational waves  penetrate all of cosmic history at sub-Planckian densities, and even propagate some distance into extra dimensions, these techniques will access directly many phenomena that are currently hidden from view. 

This brief survey represents an introduction and  update; more detail on many of the ideas presented here appear in earlier reviews and surveys\cite{Hogan:1998dv,Maggiore:1999vm,Cutler:2002me,Hughes:2002yy,Buonanno:2003th,Chongchitnan:2006pe}. 
Also, since the emphasis is on classically-generated waves that might be directly detected in the near future (and not on cosmic background techniques), inflationary quantum perturbations are little discussed, even though they dominate calculations in the cosmological literature;  direct detection of these waves might occur in the more distant future\cite{Efstathiou:2006ak,Boyle:2005se,Smith:2005mm,Crowder:2005nr}.
\section{the invisible ages of cosmic history}

An informative   plot of all  cosmic history is a ``redshifted Hubble frequency diagram'', shown in Figure 1.
The inverse apparent  horizon size at a scale factor $a$, or Hubble frequency $H= \dot a/a$, is redshifted today to a lower 
frequency,
$aH=\dot a$. We plot this observed frequency
 as a function of $a$ (in units where the current scale factor $a_0=1$).
 Several simple scalings are: 
$\Lambda$ or slow roll inflation corresponds to $\log \dot a\propto\log a$; matter-dominated evolution gives $\log \dot a\propto -(\log a)/2$; radiation-dominated evolution gives $\log \dot a\propto -\log a$. A standard inflationary history with inflation at the GUT scale is shown as the uppermost solid line. Inflation ending at much lower energy is shown as other solid lines. The upper dashed lines show a reference universe that stays permanently radiation-dominated.\cite{Bjorken:2004an} 
For reference, note  that a frequency of about 0.1 millihertz today corresponds to the Terascale--- a temperature of about 1TeV, and an apparent horizon size of about 1 mm:
\begin{equation}
f_0=\dot a(t)\approx 10^{-4}{\rm Hz}\  (H(t)\times 1{\rm mm}/c)^{1/2}\approx  10^{-4}{\rm Hz}\ (T/1{\rm TeV})
\end{equation} 
\begin{figure} \label{fig: figure1}
\epsfysize=3in 
\epsfbox{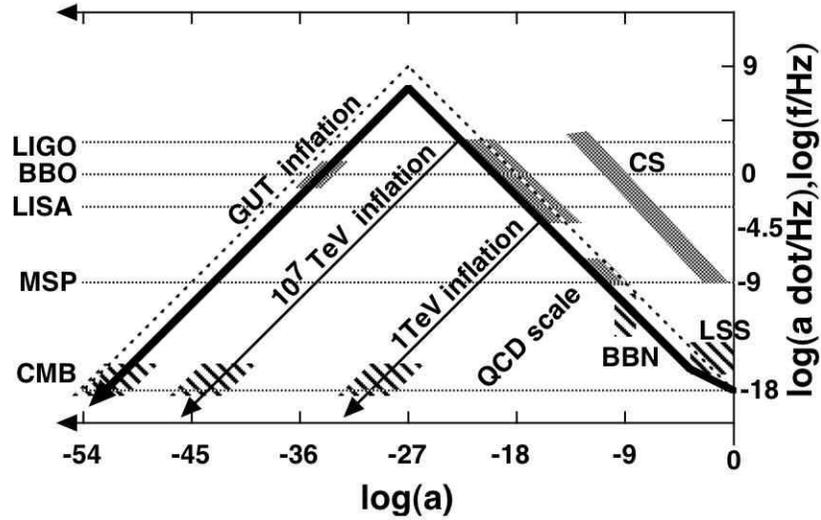} 
\caption{Reshifted Hubble frequency $\dot a$ as a function of cosmic scale factor $a$ (both base-10 log scales). Horizontal lines are shown indicating observable gravitational wave frequency bands: the current Hubble frequency (CMB techniques),   inverse years to  decades (millisecond pulsar timing, MSP), millihertz (LISA), one Hz ( the vision mission BBO, Big Bang Observer), and kilohertz (LIGO,VIRGO,GEO). Heavy solid lines show a standard inflationary model that has expanded by about 27 orders of magnitude since GUT scale inflation.  Later-ending inflationary models are also shown, reheating at $10^7$ TeV scale energy ($f\approx 10^3$ Hz,$H\approx 10^{-25.5}$ sec)  and 1 TeV scale ($f\approx 10^{-4}$ Hz, $H\approx 10^{-11.5}$ sec).  Shaded regions correspond to accessible cosmological sources of gravitational waves: on the right branch, classical sources such as defects, phase transitions, and brane displacement excitations; on the left branch, inflationary quantum sources. Shaded region above the main lines shows the scale factor where  light cosmic superstrings (CS) emit most of their observed background energy at each frequency. Hatched regions show the range covered by currently accessible cosmological data on CMB, large scale structure (LSS) and big bang nucleosynthesis (BBN).}
\end{figure}

The cross-hatched regions show areas about which we currently have at least some fairly direct data. At the lower right, we have the universe since nucleosynthesis; at the lower left, standard GUT scale inflation mapping onto the CMB anisotropies. 
The diagram has the virtue of expanding much intermediate history over a wide range of scales. The figure can be regarded  as applying to extra dimensions as well, very early when these may be in play, as long as they can be described by a single scale factor.

 Horizontal lines in this plot show constant observed frequency; the universe lines cross these at observed Hubble frequencies in a given model; above this corresponds to frequencies within the horizon (that is, less than $H$) at a given $a$. Lines are shown corresponding to  frequency bands for ground-based interferometers,  space-based interferometers, and pulsar timing techniques. 
 The shaded regions correspond to areas where directly observable cosmic gravitational wave backgrounds might be generated. Those to the right of the peak correspond to classical processes; those to the left correspond to quantum processes. 

All the current probes revealing the character of relic structure--- via CMB, large scale structure, and quasar absorption--- extend only four or five orders of magnitude up from the bottom of the graph. Constraints on cosmic history from the CMB spectrum and nucleosynthesis extend farther, back to weak decoupling. The formation of WIMP-type dark matter and neutrino decoupling happen at about the same time, and  axion dark matter condenses at the QCD epoch.  From earlier times, microscopic information tends to be thermalized, apart from a few conserved quantities; indeed, perhaps even baryon number does not survive from far above the Terascale. Events in the top two-thirds of the figure, including most of cosmic expansion history (in log units),  are currently invisible. That allows for a very wide range of possible new physics.

  Gravitational waves have the potential  to open up much of  this   unexplored region to direct observation. They will at least constrain models in a meaningful way, and may  reveal  new sources of gravitational waves.  There is a possibility of obtaining detailed information about cosmic  activity on mesoscopic scales, and about  what physics was up to during those ``invisible ages''.

\begin{figure} \label{fig: figure2}
\epsfysize=3.5in 
\epsfbox{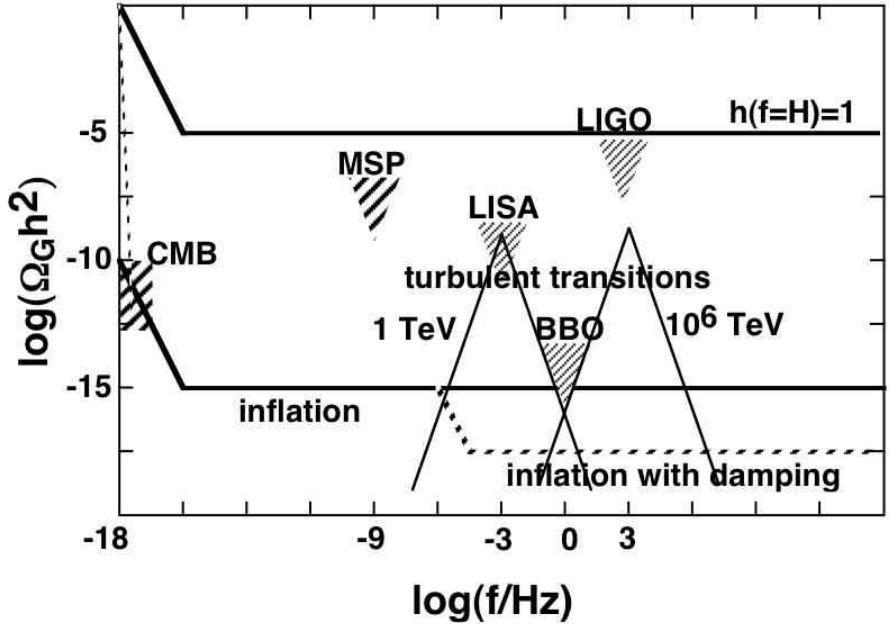} 
\caption{Overview of spectra of  gravitational  wave backgrounds from high redshift. The energy density of gravitational waves per log frequency, in units of the critical density, is shown for  inflationary quantum (tensor)  fluctuations, and  for turbulent cosmic phase transitions at 1 TeV and $10^6$ TeV (which peak somewhat above the Hubble frequencies at those times since the largest scale motions are subhorizon in scale).  Relativistic motion close to the horizon scale can create classical backgrounds that approach within a few orders of magnitude of the CMB energy density, and are far more intense  than inflationary quantum generation.  Inflation is shown here for an optimistic large amplitude and flat spectrum, but this neglects effects that might make it smaller:  a dashed line shows the sign of effects of damping or postinflationary modifications to the equation of state.  CMB shows the possible region explored by   microwave background polarization.}
\end{figure} 

\section{new classical sources of gravitational waves}
A schematic spectrum of the gravitational wave background is shown in figure 2.  An optimistic  estimate of conventional inflationary waves is shown, saturating the current upper limit on tensor/scalar ratio, little damping, and a relativistic equation of state. The current limit on the tensor/scalar ratio limits these to about $10^{-10}$ of the CMB energy density, but in many inflation models it is much less\cite{Efstathiou:2006ak};  roughly speaking, the metric perturbation on the horizon is of the order of the inflationary expansion rate in Planck units.  It is not that surprising that the backgrounds are weak, since the production process corresponds to about a single graviton quantum per Hubble 4-volume during inflation; the occupation numbers are still at best of the order of unity when they arrive at our detectors. (However, note that there are less conventional versions of inflation, particularly pre-big-bang scenarios, with quite different and sometimes quite intense predicted spectra at high frequencies, again illustrating how much there is to learn from the gravitational wave backgrounds; see for example\cite{Buonanno:2003th,Buonanno:1996xc}).

The new classical sources discussed here are potentially much stronger in certain ranges of frequency; the natural scale of energy density for these is the CMB density, times inefficiency factors which are smaller than unity by a factor depending on powers of the scale relative to the Hubble length.
Note that LISA can detect backgrounds down to  about six orders of magnitude less energy than the relativistic plasma, so it can detect sources with  net gravitational radiation efficiency as small as about $10^{-6}$. More explanation of 
various parts of this figure can be found in reviews and summaries cited above; a more detailed discussion for LISA in particular, exploiting the use of Sagnac calibration for extracting broad band backgrounds, is  in\cite{Hogan:2001jn}.

\subsection{phase transitions and relativistic turbulence}

During an early first order phase transition, the universe ``boils''-- it nucleates bubble of new phase, and the growth of bubbles converts internal energy (the latent heat of transition) to relativistic flows on the nucleation scale.  For a strong transition the separation between nucleated bubbles is macroscopically large,  leading to flows that are coherent on scales about one to two orders of magnitude smaller than  the horizon size, a dimensionless ratio which determines the characteristic peak frequency and radiation efficiency.    The flows generate modestly relativistic bulk turbulent velocities and accelerate fluid mass, leading to generation of  gravitational waves. The kinetic energy dissipates  to smaller scales in a turbulent cascade, creating a power law of higher frequency radiation.

Gravitational wave spectra were estimated for QCD\cite{witten84}  and  electroweak\cite{hogan86} phase transitions during the 1980's. These papers estimated basic parameters such the characteristic frequency and intensity of the backgrounds, which depend mainly on the critical temperature and the  degree of supercooling (determined by the latent heat of transition and the nucleation process, e.g. \cite{Steinhardt:1981ct,hogan83}). Further   work during the 1990's down to the present \cite{Kosowsky:1992rz,Kosowsky:1991ua,Kosowsky:1992vn,Kamionkowski:1993fg,Kosowsky:2001xp,Dolgov:2002ra,Apreda:2001us,Nicolis:2003tg,Grojean:2006bp,Caprini:2006jb} made more detailed inroads into accurate models of the spectra, although these are still not realistic or definitive, as one might expect given the difficulties in understanding turbulent processes. Figure 3 shows an enlarged view of the LISA band, with a typical peak frequency and amplitude shown.  Note the importance of understanding the  astrophysical foregrounds in this frequency range with similar confusion-limited spectra, especially galactic and extragalactic dwarf binaries\cite{Farmer:2003pa}; this will likely be the key limiting factor of the sensitivity to high redshift transitions
\cite{Coward:2006df}.

Progress has continued on theoretical understanding of the physics of the transition itself.\cite{ew,Grojean:2004xa}
Recent interest is especially prompted by the approach of real data on the Terascale  soon to come from CERN's Large Hadron Collider. LISA's band is well tuned to phase transitions at LHC scales  since its peak sensitivity corresponds to about a tenth of the horizon scale at 1 TeV. 
It has been  proposed that the elecroweak transition is responsible for the departure from equilibrium that brought about baryogenesis; these scenarios could be directly confirmed by a LISA detection.

\begin{figure} 
\epsfysize=3.5in 
\epsfbox{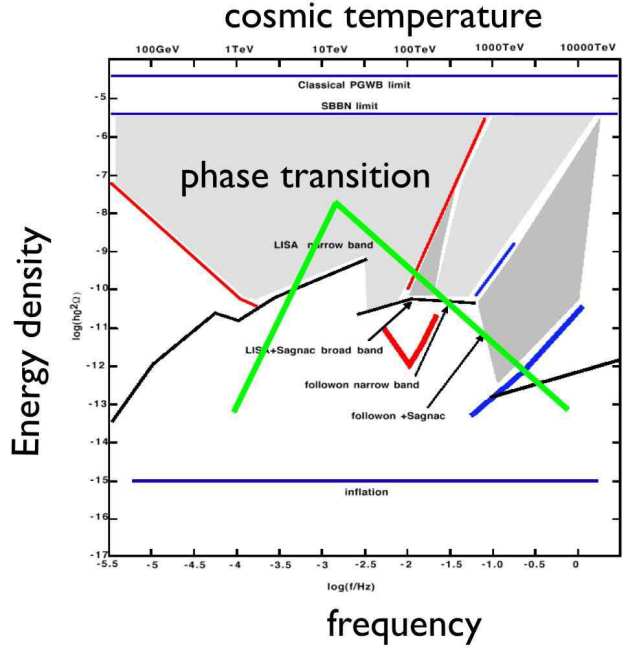} 
\caption{A schematic  background spectrum from a strong $\approx$TeV-scale phase transition,  superimposed on a   plot of LISA's sensitivity to broadband backgrounds, from\cite{Hogan:2001jn}. The upper axis shows the energy scale (that is, roughly the fourth root of the mean energy density) corresponding to  the redshifted Hubble frequency. The sample background shown is not calculated exactly but is typical of estimated spectra, showing how the shape of the spectrum falls off on each side of the characteristic frequency and  might be used as  a diagnostic of the source. The lower boundaries of the shaded regions show roughly where astrophysical backgrounds or instrument noise limit the measurements.}
\end{figure}

\subsection{dynamics of extra dimensions}

String theory requires many extra dimensions. The sizes of the dimensions, their shapes, and how they are stabilized are unknown. All we really know is that for string theory to be right, the extra dimensions must be small, or their radii of curvature must be small, so that they do not appear today in direct tests of the  gravitational inverse-square law at   submillimeter scale\cite{Hoyle:2004cw}. (The scales probed by Standard Model fields are of course much smaller than this, but they might be confined to a 3-brane living in a  larger dimensional space.)

Since it turns out the Hubble length at the Terascale is also about a millimeter, the current threshold of ignorance happens to be about the same in the laboratory gravity,  particle/field, and  cosmological realms. Thus we find lab experiments, accelerator physics, and LISA cosmology converging on the same new regime in very different ways. It is even possible that new properties of gravity on this scale are related to cosmic dark energy (whose energy density is about $(0.1{\rm mm})^{-4}$).

The formation of extra stable dimensions introduces  sources of free internal energy that might be released coherently on a macroscopic scale.\cite{Hogan:2000is,Hogan:2000aa}  Moreover it introduces another kind of mechanism for generating gravitational waves: motion and curvature of our standard model brane in the extra dimensions. 
At the time of the  localization of the graviton by curving extra dimensions, the position of our brane  might be a random variable above the horizon/curvature scale,  and in general, spatially inhomogeneous modes would be excited by the Kibble mechanism. (Note that most braneworld scenarios concentrate instead on generation of quantum noise in an already-stabilized brane system; see e.g. \cite{Jones:2002cv,Sarangi:2002yt,Jones:2003da,Kachru:2003sx,Durrer:2005dj,Firouzjahi:2005dh,Hiramatsu:2006bd})

Alternatively is also possible that the new scales are all much smaller than the Hubble length when the stabilization occurs. In this case,  the behavior of the extra dimensions can be described by a scalar order parameter as a function of 3+1D position, and  the effect in 3+1D spacetime is similar to the phase transitions just discussed.

\subsection{late inflationary reheating:  the ``loud big bang''}

There is no mandatory reason to  assign an extremely high redshift to inflation. Reheating ends as late as the Terascale in many scenarios (particularly those braneworlds where the Planck scale is not far above the Terascale), and could have ended  as late as the QCD epoch or even electroweak decoupling  without directly affecting current cosmological  observables. If we are fortunate, it might have ended with enough classical gravitational  noise  to be accessible to direct gravitational wave detection. 

There is also no fundamental reason to assume a very quiet reheating.  Inflation itself is an extraordinarily coherent behavior of a scalar field; reheating is a process that eventually converts its  internal potential energy into a thermal mix of relativistic particles. In many scenarios (especially ``hybrid'' ones), the conversion begins with macroscopically coherent but inhomogeneous motions   that eventually cascade to microscopic scales. Quantum coherent processes such as ``preheating'' transform into coherent classical motions which, like the phase transitions discussed above, generate backgrounds of the order of $10^{-3}$ of the thermal plasma density\cite{Kofman:2005yz,Felder:2006cc,Easther:2006gt}. (As with those transitions, the frequency of the background only matches the gravitational wave detectors if the final activity of this type occurs well below the GUT scale).  
A closely related effect is the formation of nonlinear horizon-scale domains in axion- or Higgs- like fields of very low mass; in this case the field's internal energy is converted not into radiation, but  into miniclusters of cosmic dark matter\cite{Zurek:2006sy}. 

The spectra from the above three sources resemble each other: a broad peak around a characteristic frequency,   a power law tail at high frequency from the cascade, and a steep rolloff at low frequency from causality  (the redshifted Hubble frequency), since low frequencies only come from small velocity flows which are inefficient radiators of gravitational waves.  The above examples show that LISA is positioned to detect direct evidence of first-order phase transitions, or indeed any significant sharing of internal energy in sub-horizon-scale bulk motions  near and somewhat above the Terascale.

\subsection{backgrounds and bursts from cosmic superstrings}

Cosmic strings have been studied for many years as a possible new form of mass-energy with new and distinctive astrophysical effects\cite{kibble,Zeldovich:1980gh,Vilenkin:1981iu,Vilenkin1981,Hindmarsh:1994re,vilenkinshellard} They were originally  conceived in  field theory as defects caused by broken U(1) symmetries in Yang-Mills theories.  Recently they have re-emerged as  possible quasi-stable structures of  fundamental string theory, sometimes called cosmic superstrings  \cite{Copeland:2003bj,Jackson:2004zg,polchinski,Polchinski:2004hb,davis,kibble2,vilenkin2} that tend to  arise naturally as U(1) symmetries are broken in models of brane  inflation \cite{Jones:2002cv,Sarangi:2002yt,Jones:2003da,Kachru:2003sx}.

A tangled  net  of superstrings forms by a process of Kibble quenching after cosmological inflation.  The primordial net of long strings continually intersects with itself, spawning isolated, oscillating loops that ultimately radiate almost all of their energy into 
gravitational waves\cite{Vilenkin1981,Hogan:1984is,Vachaspati:1986cc,Bennett:1989yp,Caldwell:1991jj,Caldwell:1996en}.   
Although the fundamental physics differs widely for different types of strings, their quantitative gravitational effects are mainly governed by one fundamental parameter,  the dimensionless mass per length or tension $G\mu$ (in Planck units with $G=m_{Planck}^{-2}$). Current limits on gravitational wave backgrounds (from pulsar timing) already suggest that if  superstrings exist, they must be so light that they have no observable astrophysical effects other than their gravitational radiation.  

Figure 4 shows   predicted stochastic background spectra\cite{Hogan:2006we}
 from strings  for various values of $G\mu$. The current pulsar limits corresponding to $G\mu\approx 10^{-10}$ already  significantly constrain brane cosmology, and LISA will probe beyond this limit by several orders of magnitude, to $G\mu\approx 10^{-15}$. The spectrum from superstrings is clearly distinguishably different from that of phase transitions or any other predicted source: it is nearly flat (in $\Omega$ units) over many decades at high frequencies, including the range where LISA is likely to observe it.
 
 There  is a possibility, if the strings are not too much  lighter than current limits, that occasional distinctive bursts might be seen from nearby loops that happen to  beam gravitational waves in our direction from cusp catastrophes in the loop trajectory\cite{Hogan:2006we,Damour:2004kw,Siemens:2006vk}.  These rare events, if they are intense enough to stand out above the background,   are recognizable  in principle from their  universal waveform, which derives just from the geometry of the cusps.
Approaching time $\delta t$ from the moment of the catastrophe,  in units  given by the fundamental mode of the loop, the metric strain amplitude due to  radiation beamed from a cusp varies like  $h(\delta t)\propto \delta t^{1/3}$, and is beamed within an angle $\theta\propto\delta t^{1/3}$.  (That is, if observed at angle $\theta$ from the beam direction, the cusp waveform behavior is smoothed out at $\delta t< \theta^3$.) 
Although individual burst events, if detected,  give the clearest signature of a string source,   the  first detectable sign of a superstring loop population is likely their integrated confusion-limited stochastic background\cite{Hogan:2006we}.

While LISA's upper limits will certainly provide interesting constraints and eliminate classes of cosmological theories, the actual discovery of an identifiable superstring background  (and even better,  possible but less likely, superstring bursts)  would be  a direct observation of a    stringy signature in nature.   Measured properties would provide insight into physics beyond field theory and classical general relativity, and into the physics underlying cosmic inflation.  Real  data would  go a long way  to help shape the rich mathematical insights of string theory into a model of the real world.
 
\begin{figure} 
\epsfysize=3in 
\epsfbox{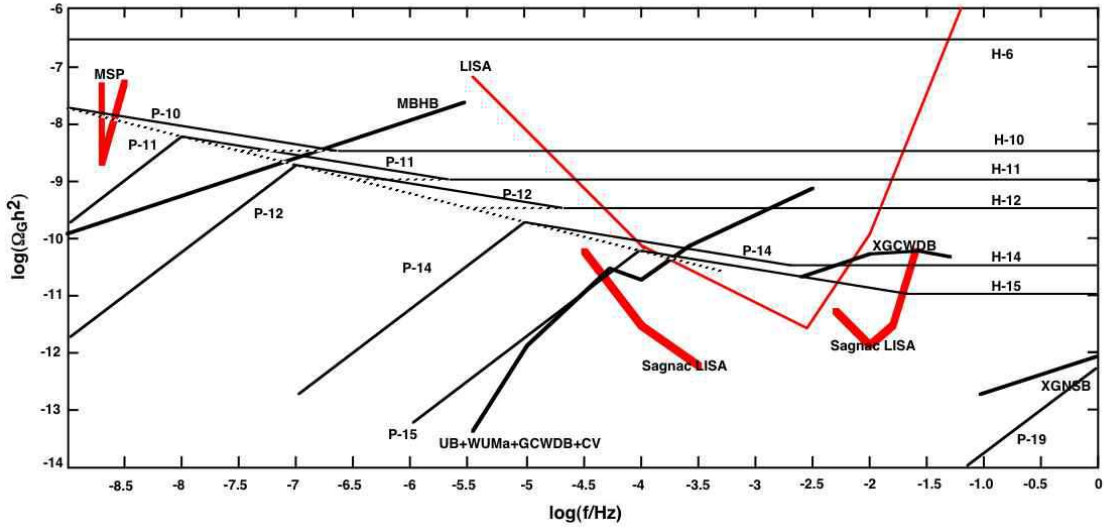} 
\caption{Predicted gravitational wave backgrounds from a population of superstring loops,  and  a summary of foreground and instrument noise sources\cite{Hogan:2006we}. Broad band energy density is shown in units of the critical density for $h_0=1$, as a function of frequency, for  a model where the typical size of newly formed loops is a fraction $\alpha=0.1$ of the horizon. Radiation from   loop populations at  high redshift (H)  and present-day (P) is shown, labled by the value of $\log_{10}(G\mu)$. Noise levels are shown for current millisecond pulsar data (MSP), and the projected  LISA sensitivity in maximum resolution and Sagnac modes. Confusion noise is shown for massive black hole binaries (MBHB), the summed Galactic binary population including binary white dwarfs (UB+WUMa+GCWDB+CV), and extragalactic populations of white dwarfs (XGCWDB) and neutron stars (XGNSB).  Dotted curves show the  contributions of $z>1$ loops where they are subdominant to the P contributions. Current (MSP) sensitivity is at about $G\mu\approx 10^{-10}$, and LISA will reach to around $G\mu\approx 10^{-15}$.}
\end{figure}

\begin{theacknowledgments}
This work was supported by NSF grant AST-0098557 at the University of
Washington. I am grateful to J. D. Bjorken for helpful discussions.
\end{theacknowledgments}

\end{document}